\documentclass{iau}
\usepackage{graphicx,natbib}
 
\newcommand{\apj}{ApJ}           

\newcommand{\aap}{A\&A}

\title{A 3D view of the Hydra I galaxy cluster core –- I. Kinematic substructures}

\author[Hilker et al.]{Michael Hilker$^1$, Carlos Eduardo Barbosa$^{1,2}$, Tom Richtler$^3$, Lodovico Coccato$^1$, Magda Arnaboldi$^1$ \and Claudia Mendes
de Oliveira$^2$}

\affiliation{$^1$European Southern Observatory, Garching, Germany,
email: {\tt mhilker@eso.org} \\
$^2$Universidade de Sao Paulo, Sao Paulo, Brazil \\
$^3$Universidad de Concepci\'on, Concepci\'on, Chile \\}

\pubyear{2014}
\volume{309}
\jname{Galaxies in 3D across the Universe}
\editors{B. L. Ziegler, F. Combes, H. Dannerbauer, M. Verdugo, eds.}

\begin{document}

\maketitle

\begin{abstract}
We used FORS2 in MXU mode to mimic a coarse 'IFU' in order to measure the
3D large-scale kinematics around the central Hydra\,I cluster galaxy NGC\,3311.
Our data show that the velocity dispersion field varies as a function of radius and
azimuthal angle and violates point symmetry. Also, the velocity
field shows similar dependence, hence the stellar halo of NGC\,3311 is a
dynamically young structure. The kinematic irregularities coincide in position
with a displaced diffuse halo North-East of NGC\,3311 and with tidal features of
a group of disrupting dwarf galaxies. This suggests that the superposition of
different velocity components is responsible for the kinematic substructure
in the Hydra\,I cluster core.
\end{abstract}

\firstsection
\section{Introduction, target and method}

The mass determination of central cluster galaxies based on kinematical data 
normally uses simplified assumptions of virial equilibrium and spherical 
symmetry. The round appearance of these galaxies and the smooth distribution 
of hot X-ray emitting gas might suggest that these assumptions are justified.
On the other hand, it has been shown that the halos of massive ellipticals
grow by a factor of about 4 in mass since z$=$2 (van Dokkum et al. 2010).
The mass growth is dominated by the accretion of low mass systems (minor
mergers). Thus, one might expect that accretion events leave kinematical
signatures in the phase space of the outer stellar population especially of
central cluster galaxies.

Here we present the vivid case of the central giant elliptical of the Hydra\,I 
cluster, NGC\,3311. This early-type galaxy dominated cluster is regarded as
dynamically evolved. Recent photometric and kinematical studies of the diffuse
stellar light, planetary nebulae and globular clusters in the core of Hydra\,I,
however, have shown that 1) NGC\,3311 exhibits a steeply rising velocity 
dispersion profile (Ventimiglia et al. 2010, Richtler et al. 2011), 2) the velocity
dispersion profiles differ from each other in different azimuthal directions, as
judged from longslit analyses (Ventimiglia et al. 2011, Richtler et al. 2011),
and 3) the diffuse light is not centered around NGC\,3311's main spheroid,
but it is displaced towards the North-East by about 15 kpc (Arnaboldi et al. 2012). 

In order to find kinematic signatures of these substructures and
to disentangle the past and present active assembly history of the Hydra\,I 
cluster core, we used FORS2 in MXU mode (ESO programme 088.B-0448, PI:
T.\,Richtler) to mimic a coarse `IFU'. Our novel 
approach is to place short slits in an onion shell-like pattern around 
NGC\,3311 to measure its 3D large scale kinematics out to 3 effective radii.
Sky slits were positioned far outside of NGC\,3311's main body and bright halo.
The borders of the `IFU' spaxels were defined via Voronoi tesselation to create
kinematic maps (see Fig.\,1). The S/N of our final spectra ranges from $>$20 to 2 from
the inner to the outer radii.

\section{Results and conclusions}

\begin{figure}
\centering
\includegraphics[width=0.7\columnwidth]{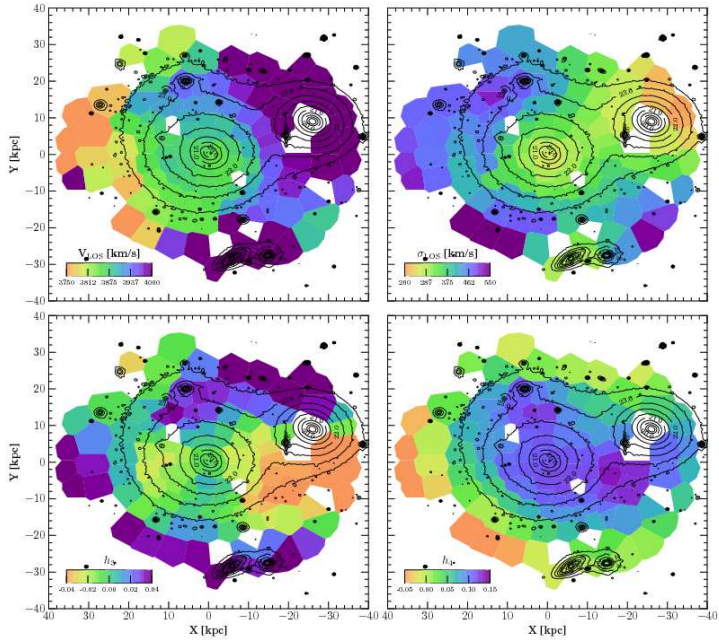} 
\caption{{\bf Upper left:} Radial velocity map. {\bf Upper right:} Velocity dispersion map.
{\bf Lower left:} Skewness (h3) map. {\bf Lower left:} Kurtosis (h4) map. All maps were
smoothed with a locally weighted scatterplot smoothing algorithm (Cleveland 1979).
The contours are surface brightness levels of NGC\,3311 (in the center) and
NGC\,3309 (on the right) $V$-band light between 19 and 23.5 mag/arcsec$^2$ in
steps of 0.5 mag.}
\end{figure}

The main results of our analysis, shown in Fig.\,1, are:
1) there are pronounced azimuthal variations both in radial velocity and velocity
dispersion as a function of galactocentric distance, explaining the above mentioned discrepancies from
previous longslit results; 2) in the North-East there are significant small scale
variations in both radial velocity and velocity dispersion data points; 3) the very large
velocity dispersion values in some parts of the cluster core ($\sigma>500$ km/s)
probably point to a superposition of kinematical substructures; and 4) the
displaced diffuse stellar halo around NGC\,3311 coincides with regions
of positive h3 and h4 values,
also evidence for more than one velocity component in the stellar halo.

We conclude that the stellar halo around NGC\,3311 in the core of Hydra\,I is
still forming and is not in dynamical equilibrium. Probably, a group of infalling
dwarf galaxies and their tails are responsible for the kinematical substructures.
The general lesson is that one has to take care when inferring the properties of
the central dark matter halo around NGC\,3311 from kinematical data.

%

\end{document}